\documentclass[conference]{IEEEtran}
\IEEEoverridecommandlockouts
\usepackage{cite}
\usepackage{amsmath,amssymb,amsfonts}
\usepackage{algorithmic}
\usepackage{graphicx}
\usepackage{textcomp}
\usepackage{xcolor}
\usepackage[utf8]{inputenc}
\usepackage{orcidlink}
\usepackage{amsmath,amsfonts}
\usepackage{algorithm}
\usepackage{array}
\usepackage[caption=false,font=normalsize,labelfont=sf,textfont=sf]{subfig}
\usepackage{stfloats}
\usepackage{url}
\usepackage{verbatim}
\usepackage{afterpage}
\usepackage{float}
\def\BibTeX{{\rm B\kern-.05em{\sc i\kern-.025em b}\kern-.08em
    T\kern-.1667em\lower.7ex\hbox{E}\kern-.125emX}}
\begin{document}

\title{\textbf{Bio-RV}: Low-Power Resource-Efficient RISC-V Processor for Biomedical Applications

\thanks{This work was supported by the Special Manpower Development Program for Chip to Start-Up (SMDP-C2S), the Ministry of Electronics and Information Tech(MeitY), and the MeitY), Government of India, Grant: EE-9/2/21 - R\&D-E.

\hspace*{2cm}979-8-3315-4970-1/26\$31.00~\copyright~2026 IEEE}
}

\author{
    \IEEEauthorblockN{{Vijay Pratap Sharma\IEEEauthorrefmark{1}\orcidlink{0009-0008-1838-5179}},
    Annu Kumar\IEEEauthorrefmark{1}\orcidlink{0009-0009-2197-3745},  Mohd Faisal Khan\IEEEauthorrefmark{1}\orcidlink{0009-0008-2235-8341},\\
    Mukul Lokhande\IEEEauthorrefmark{1}\orcidlink{0009-0001-8903-5159}, Member, IEEE,
    Santosh Kumar Vishvakarma\IEEEauthorrefmark{1}\orcidlink{0000-0003-4223-0077}, Senior Member, IEEE.}
    \IEEEauthorblockA{\IEEEauthorrefmark{1}NSDCS Research Group, Indian Institute of Technology Indore, India}
    Email: skvishvakarma@iiti.ac.in \textbf{(Corresponding Author)}
}

\maketitle
\begin{abstract}

This work presents Bio-RV, a compact and resource-efficient RISC-V processor intended for biomedical control applications, such as accelerator-based biomedical SoCs and implantable pacemaker systems. The proposed Bio-RV is a multi-cycle RV32I core that provides explicit execution control and external instruction loading with capabilities that enable controlled firmware deployment, ASIC bring-up, and post-silicon testing. In addition to coordinating accelerator configuration and data transmission in heterogeneous systems, Bio-RV is designed to function as a lightweight host controller, handling interfaces with pacing, sensing, electrogram (EGM), telemetry, and battery management modules. With 708 LUTs and 235 flip-flops on FPGA prototypes, Bio-RV, implemented in a 180 nm CMOS technology, operates at 50 MHz and features compact hardware footprint. According to post-layout results, the proposed architectural decisions align with minimal energy use. Ultimately, Bio-RV prioritises deterministic execution, minimal hardware complexity, and integration flexibility over peak computing speed to meet the demands of ultra-low-power, safety-critical biomedical systems.

\end{abstract}


\section{Introduction}

RISC-V has emerged as an open-source instruction set architecture (ISA), with applications ranging from high-performance computing (HPC) data centres to ultra-low-power embedded and Internet of Things (IoT) systems. Its open ecosystem and absence of licensing constraints make it attractive to application-specific systems-on-chip (SoCs) \cite{ref4, flex-pe}, primarily due to the costly market traditionally dominated by proprietary ISAs, such as Intel's x86 for HPC and ARM's ISA for TinyML systems. In recent years, RISC-V has been increasingly considered for safety- and reliability-critical embedded applications, such as biomedical and implantable devices\cite{ref10, ref12, ref0, ref00}.

Implantable biomedical devices, such as pacemakers, impose specific requirements on their host processors that differ from those of conventional computing systems intended for widespread use or high performance. These systems operate under extremely low duty cycles, execute deterministic control workloads, and must support post-silicon testability, firmware validation, and long-term reliability over the product lifetimes spanning multiple years. In such environments, peak throughput and instruction-level parallelism are far less critical than predictable execution, minimal silicon footprint, ultra-low power consumption, and controllable execution behaviour. Most existing RISC-V processors are optimised either for performance (e.g., BOOM, CVA6, SweRV) or for general low-power embedded use (e.g., Ibex, PicoRV32)\cite{ref1, ref2,ref3, ref5, ref6, ref7, ref8, ref9, ref10, ref11, ref12}. High-performance cores rely on deep pipelines, speculation, virtual memory, and complex control logic, resulting in increased size and energy consumption. Conversely, numerous lightweight RISC-V cores emphasise instruction throughput or FPGA efficiency, yet offer minimal assistance for explicit execution control, external observability, and post-fabrication programmability, which are crucial for the deployment and certification of medical-grade ASICs.


\begin{figure*}[!t]
    \centering
    \includegraphics[width=0.85\textwidth]{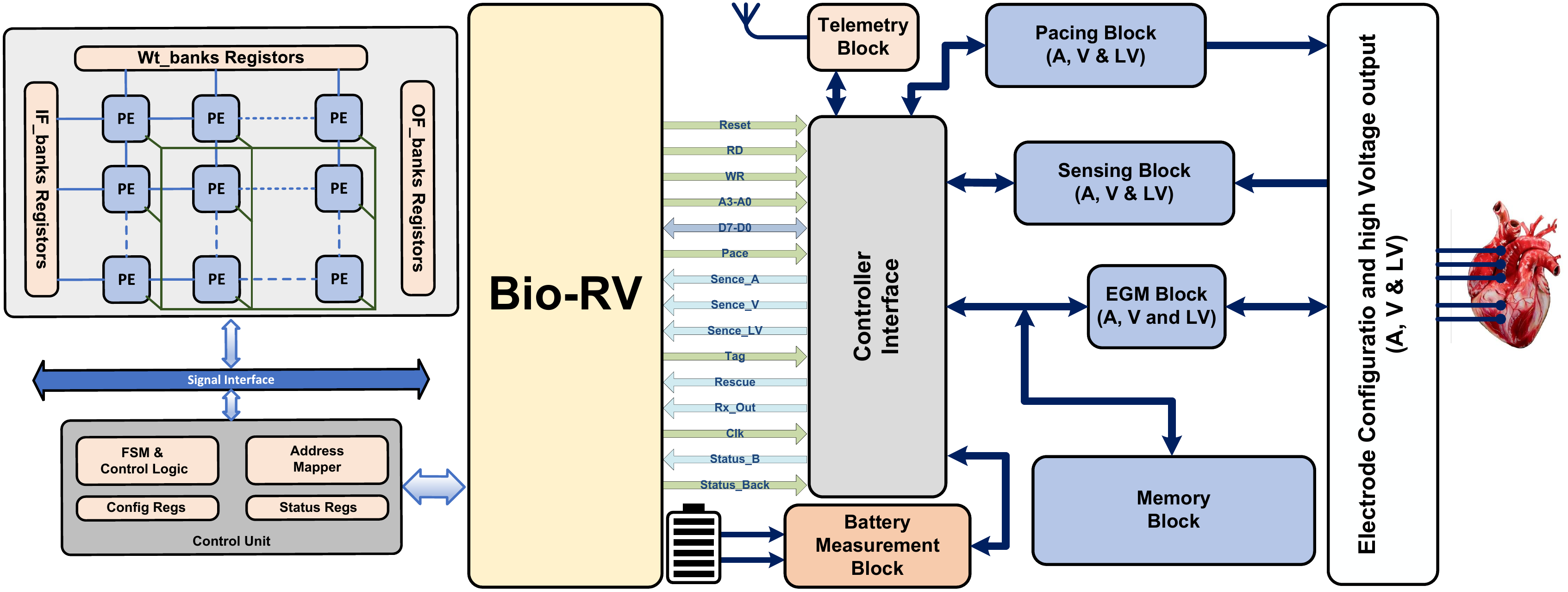}
    \caption{
    Typical TinyML SoC architecture for biomedical iPacE-CHIP applications, highlighting the Bio-RV host processor and interfaced accelerator architecture and pacing, sensing, electrogram (EGM), telemetry, and battery management blocks.}
    \label{risc-v:general}
\end{figure*}

In implantable pacemaker SoCs \cite{siwa, ref0, ref00, ref000}, the host processor primarily functions as an always-on control unit, coordinating pacing, sensing, telemetry, memory access, and battery management. These tasks are characterised by simple control flow, low instruction-level parallelism, and strict power constraints. Moreover, medical ASICs require features to halt execution, introduce instructions, and monitor memory status without relying on clock gating or intricate debugging systems, especially during silicon bring-up and extended field verification. Inspired by these constraints, this research presents Bio-RV, a streamlined, multi-cycle RV32I RISC-V processor designed explicitly for ultra-low-power biomedical control applications. Instead of targeting performance enhancement, Bio-RV promotes architectural simplicity, minimal resource utilisation, and intuitive execution control, enabling reliable operation and testability in implantable medical devices. A key feature of Bio-RV is its ability to handle the loading and execution of external instructions, as well as control for turning them on or off, which supports firmware programming, monitoring, and validation without unintentional execution, an essential requirement for ASIC deployment and post-silicon testing in safety-critical domains. 

In addition to pacemaker-class control workloads, Bio-RV can operate as a lightweight host processor in biomedical SoCs that contain domain-specific accelerators, such as TinyML or signal-processing engines. TinyML Biomedical Vision accelerator SoC\cite{flex-pe} includes a RISC-V host processor coupled with a systolic array and dedicated on-chip memory banks to accelerate deep neural network (DNN) computations as depicted in Fig. \ref{risc-v:general}. In such systems, the processor is not responsible for compute-intensive workloads, but instead oversees configuration, scheduling, data transportation, and control of accelerators. A control unit governs PE operation, sequencing, and data synchronisation, consisting of a finite state machine (FSM) with control logic, configuration registers, and an address mapper, and status registers. The RISC-V core programs the control unit and manages data transfers over a high-throughput AXI interface, enabling flexible mapping of various DNN layers. The proposed design is consequently positioned as a controller-oriented RISC-V core, rather than a throughput-optimised or compute-centric CPU. This study focuses on the micro-architectural design, execution, and assessment of Bio-RV, demonstrating how a thoughtfully constrained multi-cycle RISC-V architecture can achieve very low power and area while ensuring adequate flexibility for biomedical ASIC applications, specifically for pacemakers.

With the motivation described, we proposed a Bio-RV: a RISC-V processor tailored for biomedical applications, specifically for pacemakers. FPGA and post-layout ASIC outcomes in a 180 nm CMOS technology confirm the proposed method and emphasise its appropriateness for resource-limited, continuously operational biomedical systems. The key contributions are summarised as follows:

\begin{itemize}
    \item \textbf{Explicit Execution Control and External Instruction Loading for Biomedical ASICs:} 
        We propose a multi-cycle RV32I RISC-V-based Bio-RV processor that supports explicit execution turn-on/off control, as well as external instruction loading. These specifications are crucial for post-silicon testing, firmware validation, and long-term reliability assessment in implantable biomedical devices, where controlled execution and observability are mandatory.
        
    \item \textbf{Ultra-Minimal Controller-Oriented Microarchitecture Optimised for Always-On Biomedical Workloads:}
        The proposed Bio-RV employs a deliberately simple non-pipelined, multi-cycle architecture that prioritises deterministic control flow, reduced switching activity, and minimal hardware state rather than maximising instruction throughput. This makes Bio-RV well-suited for always-on control tasks in pacemaker-class systems, where predictability, energy efficiency, and silicon expense are more critical than maximum performance. 

    \item \textbf{Portable Host-Controller for Accelerator-Based Biomedical SoCs:} 
    The proposed Bio-RV is designed as a lightweight host controller to manage configuration, scheduling, and memory coordination for biomedical SoCs that incorporate domain-specific accelerators, including TinyML or signal-processing units. This FPGA–ASIC–compatible design enables seamless prototyping and implementation across various platforms, while post-layout outcomes in a 180 nm CMOS process validates its suitability for safety-critical, resource-constrained biomedical applications.
\end{itemize}

This article is organized as follows: Section~II describes the BIO-RV architecture; Section~III presents the performance evaluation; and Section~IV concludes the paper.

\begin{figure*}[!t]
    \centering
    \includegraphics[width=0.85\textwidth, height=85mm]{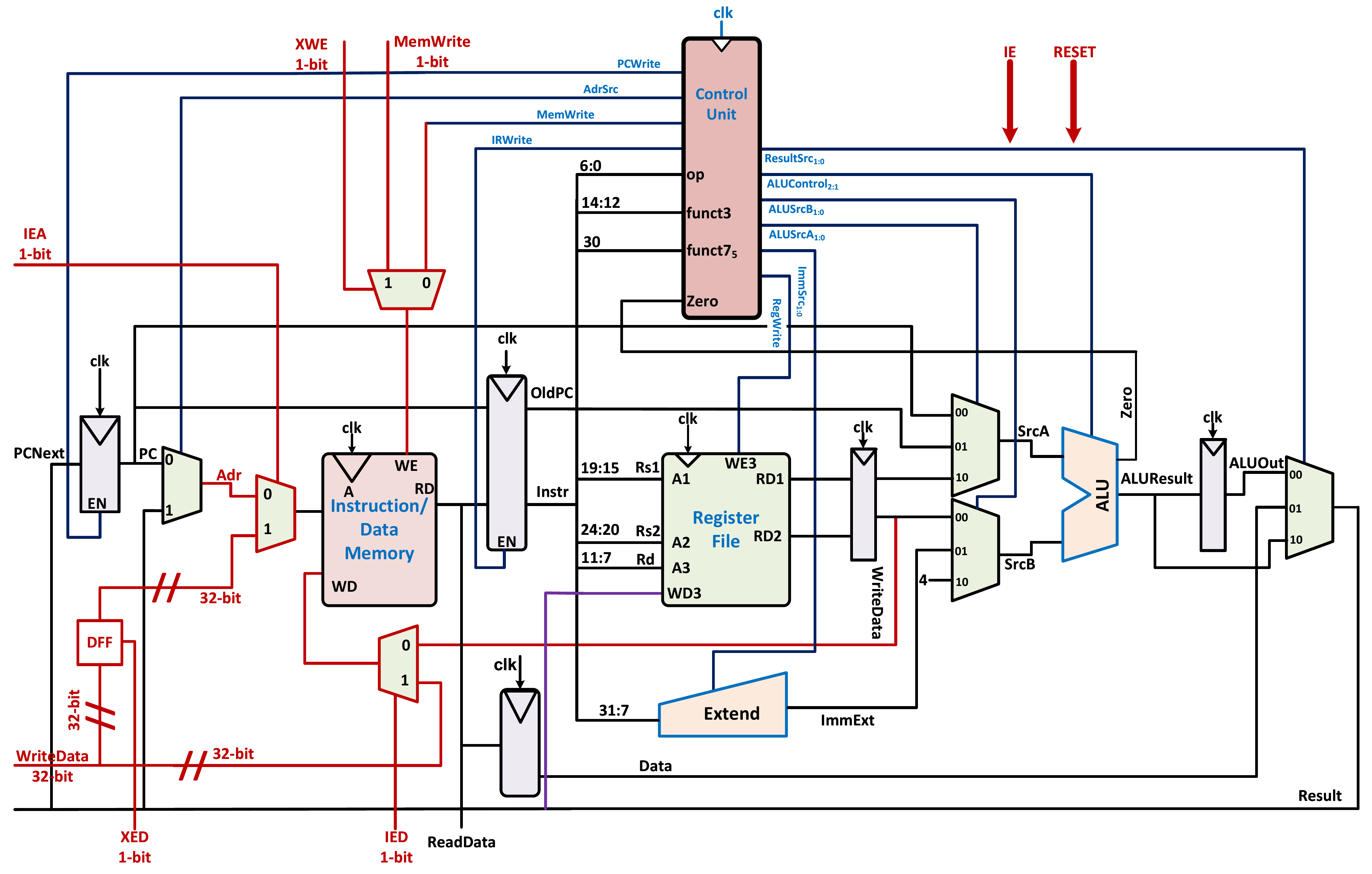}
    \caption{Detailed Data-path for the proposed Bio-RV processor}
    \label{fig:biorv}
\end{figure*}

\section{BIO-RV architecture}
 \subsection{Micro-architectural Overview}
The proposed Bio-RV, depicted in Fig. \ref{fig:biorv}, is a controller-oriented, multi-cycle RV32I RISC-V processor designed for ultra-low power and safety-critical applications. In contrast to pipelined or throughput-optimised RISC-V cores, Bio-RV places a higher priority on explicit controllability, minimum hardware complexity, and deterministic execution- all of which are critical for medical-grade ASIC deployment and always-on biomedical control workloads.

\subsection{External Instruction Loading and Explicit Execution Control}
The non-pipelined, multi-cycle architecture Bio-RV supports a basic instruction subset, including R-type, I-type, load (lw), store (sw), jump (jal), and branch (beq) instructions. The CPU uses a single shared memory for both data and instructions in accordance with the Von Neumann memory organisation. By removing redundant memory and extra address-generation logic, this design decision lowers silicon area and switching activity at the expense of more cycles per instruction- a reasonable trade-off for low-duty-cycle biomedical control tasks. The ability to load external instructions with explicit execution control is a fundamental aspect of Bio-RV's architecture. Instructions and data can be written directly into the internal instruction/data memory through the processor's 32-bit external WriteData interface. An externally accessible Instruction Enable (IE) signal is incorporated into the control logic to stop unwanted execution during programming. During instruction loading, IE is held low (IE = 0), and reset is kept low to halt program execution. Instructions can then be written successfully to memory by specifying the address and corresponding machine code. Program execution is initiated by asserting reset high to set the program counter to zero, then deasserting reset and setting IE high (IE = 1), allowing the core to begin execution from the next clock cycle. Without the need for intricate debug hardware or clock gating, deterministic firmware deployment, memory observability, and controlled start-up and shut-down operations are made possible by this clear division between programming mode and execution mode. For implanted biomedical ASICs, such features are especially crucial for long-term durability testing, field diagnostics, and post-silicon validation.
\subsection{FPGA–ASIC Friendly Implementation}
Without the use of vendor-dependent primitives, DSP blocks, or FPGA-specific memories, the Bio-RV architecture is constructed using pure RTL logic. Only a limited number of user-accessible I/O signals are exposed by the design, aside from the clock and reset, for programming and execution control. Bio-RV is appropriate for early validation and deployment in biomedical SoCs because of its simplicity, which permits easy FPGA development and a smooth transition to ASIC implementations in established technology nodes. 

\begin{table}[!t]
\caption{Comparison with SoTA RISC-V processors FPGA Utilisation}
\label{tab:fpga_comp}
\renewcommand{\arraystretch}{1.15}
\resizebox{\columnwidth}{!}{%
\begin{tabular}{|c|c|c|c|c|c|}
\hline
\textbf{Design} & \textbf{Op. Freq (MHz)} & \textbf{LUT} & \textbf{FF} & \textbf{DSP} & \textbf{Dy. Power (mW)} \\ \hline
BOOM\cite{Celio:EECS-2015-167} & 34.5 & 49865 & 25025 & 40 & 188 \\ \hline
Rocket\cite{Asanović:EECS-2016-17} & 90.9 & 5073 & 2008 & 4 & 92 \\ \hline
M RISC-V \cite{Mriscv}& 111 & 1893 & 923 & 0 & 32 \\ \hline
Zero-RIC5Y \cite{zeroRiscy}& 14.78 & 3171 & 11928 & 1 & 20 \\ \hline
HAMSA-DI \cite{ref4}& 93.67 & 6874 & 1826 & 4 & 45 \\ \hline
Shakti-E\cite{SHAKTI} & 70 & 4548 & 1813 & 4 & 100 \\ \hline
CV32E40P \cite{ref5}& 78.55 & 9072 & 2553 & 7 & 49 \\ \hline
Proposed & 50 & 708 & 235 & 0 & 15 \\ \hline
\end{tabular}}
\end{table}

This design has been sent for tapeout using SCL's 180 nm CMOS technology, confirming the proposed architecture at an established process node frequently utilised for mixed-signal and biomedical ASICs. The external WriteData interface facilitates controlled firmware deployment during silicon bring-up and testing by allowing direct instruction and data loading into the on-chip unified instruction/data memory. The output can instantly reflect the information at the specified place since read operations are asynchronous, while memory writes are synchronous. The memory can be kept in a read-only observation mode, allowing safe internal state inspection without accidental execution, by setting the execution-control signals accordingly (e.g., enabling instruction access while disabling memory writes). The Bio-RV processor has a multi-cycle execution architecture, requiring five cycles for load instructions, three cycles for branch instructions, and four cycles for arithmetic, jump, and store operations. By sharing a single memory for data and instructions, and eliminating discrete adders, this method significantly reduces hardware costs compared to a single-cycle design, albeit at the expense of more temporary registers and multiplexing circuitry. This deliberate trade-off aligns with the intended workload characteristics of biomedical control applications, where instruction throughput is not as crucial as minimising switching activity and reducing silicon space. Crucially, Bio-RV offers explicit execution control, unlike traditional multi-cycle RISC-V reference designs, where instruction memory is handled as an external block and execution cannot be halted without halting the clock. Because the same clock simultaneously advances the processor state and modifies memory contents in baseline systems, dynamic instruction loading and controlled stopping are not possible, making instruction writes during execution risky. On the other hand, Bio-RV allows for deterministic execution start/stop, memory observation, and instruction loading by separating the programming and execution phases through specific execution-control signals. Because of this feature, the proposed processor well-suited for long-term reliability testing, post-silicon validation, and ASIC deployment in safety-critical biomedical systems where controlled execution behaviour is crucial.

\begin{table*}[!t]
\centering
\caption{COMPARISON AGAINST OTHER RISC-V CORES, BASED ON THE POST-PLACE AND-ROUTE AREA AND TIMING REPORTS}
\label{tab:pnrbiorv}
\renewcommand{\arraystretch}{1.15}
\resizebox{0.85\textwidth}{!}{%
\begin{tabular}{|c|c|c|c|c|c|c|}
\hline
\textbf{Core} & \textbf{Siwa\cite{siwa}} & \textbf{Mriscv\cite{Mriscv}} & \textbf{Riscy}\cite{Riscy} & \textbf{Zero-Riscy\cite{zeroRiscy}} & \textbf{Micro-Riscy\cite{zeroRiscy}} & \textbf{Bio-RV (Proposed)} \\ \hline
Technology & 180 nm & 130 nm & 65 nm & 65 nm & 65 nm & 180 nm \\ \hline
Frequency & 20 MHz & 160 MHz & ND & ND & ND & 50 MHz \\ \hline
ISA & RV32I & RV32IM & RV32IM+DSP & RV32IM & RV32E & RV32I \\ \hline
Program Memory & 8 kB & 4 kB & ND & ND & ND & 4 kb \\ \hline
Average CPI & 4 & ND & 1.27 & 1.49 & 1.49 & 5 \\ \hline
pJ/cycle & 48.31 & 850 & 63.68 & 26.12 & 23.61 & 17.18 \\ \hline
Core Area ($\mu$m\textsuperscript{2}) & 672146 & 350250.4 & 703296 & 326592 & 200448 & 697039 \\ \hline
\end{tabular}}
\end{table*}

\section{Performance Evaluation}

The proposed Bio-RV processor is compared with a sample RISC-V cores in Table~\ref{tab:fpga_comp}, with an emphasis on indicative operational parameters and logic resource utilisation. Rather than benchmarking instruction throughput or peak performance, the main goal of this comparison is to showcase Bio-RV's relative hardware footprint and architectural simplicity. Compared to complicated out-of-order cores like BOOM ($49,865$ LUTs, $25,025$ FFs) and in-order pipelined cores like Rocket ($5,073$ LUTs $2,008$ FFs), Bio-RV has a significantly reduced logic footprint on the FPGA, occupying $708$ LUTs and $235$ flip-flops. Bio-RV exhibits substantially lower utilisation, even when compared to compact embedded-class designs, such as M RISC-V ($1,893$ LUTs, $923$ FFs). Crucially, Bio-RV preserves these blocks for domain-specific accelerators and streamlines integration on heterogeneous SoCs by avoiding the use of DSP slices or vendor-specific resources. Compared to multiple pipelined embedded RISC-V cores, the proposed solution operates at a frequency of $50$ MHz. The target application domain, which prioritises deterministic control, minimal switching activity, and energy efficiency over instruction throughput is the driving force for this deliberate design decision to reduce frequency. The operating frequency is usually determined by duty-cycle requirements and power limitations, rather than computational demands for a pacemaker-class and always-on biomedical control workloads. Overall, the findings show that, at the cost of a lower peak frequency and more cycles per instruction, Bio-RV achieves a small hardware footprint with predictable execution behaviour. In deeply embedded and safety-critical biomedical applications, where silicon cost, energy efficiency, and controllability are more crucial than raw performance, this trade-off works effectively \cite{ref20, ref21, ref22}.

\begin{table*}[!t]
\centering
\caption{Comparison of the proposed Bio-RV with other MCU architectures.}
\label{tab:mcu_comp}
\renewcommand{\arraystretch}{1.45}
\resizebox{0.95\textwidth}{!}{%
\begin{tabular}{|c|c|c|c|c|c|c|c|c|}
\hline
\textbf{Core} & \textbf{Siwa\cite{siwa}} & \textbf{8051-Compatible\cite{8051}} & \textbf{Atnega328p} & \textbf{PIC16LF1823\cite{PIC16LF1823}} & \textbf{MSP430\cite{MSP430}} & \textbf{RISC-Y} & \textbf{Micro-RISC\cite{zeroRiscy}} & \textbf{Bio-RV (Proposed)} \\ \hline
Technology & 180 nm & 180 nm & ND & ND & ND & 65 & 65 & 180 nm \\ \hline
Instruction word size (bits) & 32 & ND & 16 & 8 & 16 & 32 & 32 & 32 \\ \hline
Program Memory & 8 kB & ND & 32 kB FLASH & 2 kB & 8 kB RAM & 16 kB & 32 kB & 4 kb \\ \hline
Clock frequency & 20 MHz & 13 MHz & 0-20 MHz & 31 kHz-32 MHz & 4-16 MHz & - & - & 50 MHz \\ \hline
Average CPI & 4 & ND & 1.62 & 1.1837 & $\approx$ 1 & 1.47 & 1.49 & 5 \\ \hline
pJ/cycle & 48.31 & 70.6 & 360 & 54 & 803 & 63.68 & 23.61 & - \\ \hline
Average Power ($\mu$W) & 70 @1MHz @1.8V & 918 @13 MHz & 360 @1MHz @1.8V & 54 @1MHz @1.8V & 803 @1MHz @2.2V & - & - & 859 @50MHz, @5pF \\ \hline
\end{tabular}}
\end{table*}

Table~\ref{tab:pnrbiorv} compares the proposed Bio-RV processor with several representative RISC-V cores, considering differences in technology node, operating frequency, ISA support, program memory, average CPI, energy per cycle, and core area, on a post-place-and-route basis. In terms of technology, the proposed Bio-RV and Siwa both occupy the same $180$ nm node, whereas Mriscv is implemented in $130$ nm, and other designs utilise more advanced $65$ nm processes. Despite using an older technology node, Bio-RV operates at $50$~MHz, which is $2.5\times$ faster than Siwa ($20$~MHz) and competitive given the larger transistor sizes, though slower than Mriscv ($160$~MHz) implemented in $130$~nm.
Regarding ISA support, the proposed core implements the RV32I base integer set, similar to Siwa. It is slightly less complex than designs with DSP or multiply extensions, such as Zero-Riscy RV32IM+DSP/IM. Bio-RV features four kB of program memory, the same as Mriscv, but smaller than Siwa's eight kB. From a performance perspective, the average CPI of Bio-RV is $5$, compared to $1.27$ for Riscy and $1.49$ for Zero Micro-Riscy, which achieves higher instruction throughput due to more advanced micro-architectural optimisations. This increased CPI in Bio-RV reflects its lightweight architecture, which trades off performance for power efficiency. 
In terms of energy efficiency, Bio-RV demonstrates low energy-per-cycle at $17.18$~pJ/cycle, making it energy-efficient among all listed designs. This corresponds to approximately $2.8\times$ better efficiency than Riscy ($63.68$ pJ/cycle) and $1.3\times$ better than Zero-Riscy ($26.12$ pJ/cycle). It also substantially outperforms Mriscv ($850$~pJ/cycle) and Siwa ($48.31$~pJ/cycle) highlighting significant power efficiency benefits despite prioritising compactness and low energy consumption over peak IPC.
Regarding area, Bio-RV occupies $697{,}039~\mu\text{m}^2$, which is larger than even some $65$~nm implementations such as Micro-Riscy ($200{,}448~\mu\text{m}^2$), primarily due to the coarser $180$~nm process and possible additional application-specific logic. Compared to Siwa's die area of $672{,}146~\mu\text{m}^2$, Bio-RV's area is only about $3.7\%$ larger, yet it delivers $2.4\times$ higher frequency and energy efficiency. Bio-RV is more suited for always-on control workloads than throughput-intensive applications due to its low energy per cycle, which is consistent with its small datapath, reduced switching activity, and simplified control logic. Bio-RV exhibits a similar area to Siwa, which utilises the same technology, but achieves a higher working frequency and improved controllability features designed for biomedical ASIC deployment.

\begin{figure}
    \centering
    \includegraphics[width=0.95\linewidth]{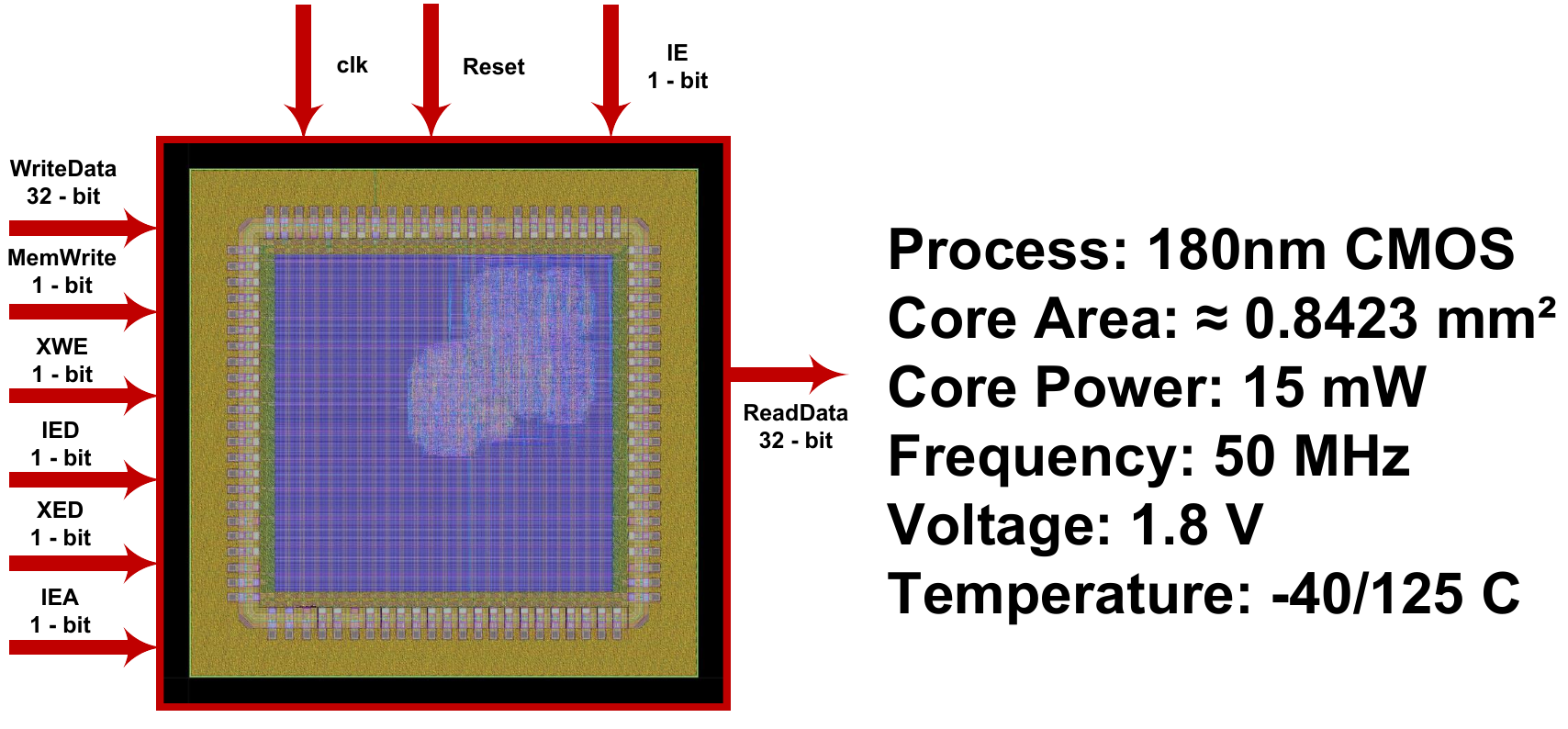}
    \caption{Final GDS-II layout of the Bio-RV implemented in 180 nm CMOS technology, including I/O pad directions.}
    \label{fig:gds}
\end{figure}

To give context on architectural scale and operating characteristics rather than a direct efficiency ranking, Table~\ref{tab:mcu_comp} compares the proposed Bio-RV processor to a variety of MCU-class systems, including older 8-bit, 16-bit, and 32-bit architectures. The presented measurements reveal significant differences between the compared systems in terms of technology node, operating frequency, execution model, and target workloads. While other contemporary 32-bit RISC-based MCUs use more sophisticated $65$ nm technology to minimise area and leakage, Bio-RV is implemented in a $180$ nm CMOS process, similar to Siwa and an 8051-compatible MCU. Compared to many ultra-low-power MCUs designed for sub-MHz or duty-cycled operation, Bio-RV operates at a higher frequency of 50 MHz, as it targets always-on control workloads. Therefore, rather than being used as direct performance comparisons, frequency and absolute power data should be understood in the context of various usage models. The proposed processor is capable of 32-bit execution, which is broader than traditional 8-bit and 16-bit architectures and comparable to contemporary RISC-based MCUs.  Although Bio-RV has a small four-kB program memory, which is in line with its intended use as a small controller for embedded and biomedical systems, where the memory footprint needs to be kept to a minimum, and firmware complexity is restricted. Because of its multi-cycle, non-pipelined architecture, Bio-RV has a higher average CPI. This design decision prioritises predictable execution and less complicated hardware over instruction performance. These trade-offs are suitable for control-dominated workloads, where maximising instructions per cycle is not as critical as execution determinism and energy efficiency. The operating frequency, voltage, activity factors, and workload assumptions that determine the energy-per-cycle and average power numbers in Table~\ref{tab:mcu_comp} differ significantly amongst the evaluated systems. In this regard, Bio-RV exhibits energy characteristics in line with its always-on execution architecture and reduced datapath. Yet, it uses more absolute power than severely duty-cycled MCUs running at far lower frequencies. All things considered, Bio-RV holds a unique position as a small 32-bit RISC-V controller installed in a mature technological node. Due to its architectural features, it is a suitable choice for embedded control and always-on biomedical applications where integration flexibility, small hardware footprint, and deterministic behaviour are more critical than ultra-low standby power or peak performance.
The physical design of proposed Bio-RV was realised with the indigenous SCL $180$ nm process technology has been submitted for fabrication through a free academic tape out initiative, as illustrated in Fig. \ref{fig:gds}.

\section{Conclusion}
This work introduces Bio-RV, a compact, biomedical-focused RISC-V processor intended for controller-level functioning in accelerator-based biomedical SoCs and implantable pacemaker control. Explicit execution control, an external instruction loading interface, and the proposed multi-cycle RV32I architecture enhances suitability for ASIC deployment and post-silicon validation in safety-critical systems, allowing for controlled firmware distribution. FPGA prototyping and post-layout analysis on a $180$ nm CMOS process shows that Bio-RV achieves low dynamic power consumption and a small logic footprint at a $50$ MHz operating frequency. The design sacrifices peak performance purposefully in favour of deterministic execution, low hardware complexity, and reduced switching activity, all of which align with the characteristics of always-on biomedical control workloads, rather than focusing on high throughput. All things considered, Bio-RV fills a specific niche as a lightweight RISC-V controller for biomedical applications, offering a valuable basis for upcoming systems that need long-term dependability, predictable behaviour, and flexible integration with established technological nodes.

\bibliographystyle{ieeetr}
\bibliography{bib}

\end{document}